# Evaluating Cross-Encoders for Semantic Similarity Assessment in Psychological Questionnaires

Isabella Kainz[1,2], Karin Labek[1], Roberto Viviani[1,3]

1 Institute of Psychology, University of Innsbruck, Innsbruck, Austria

2 University of Applied Sciences Kufstein, Tyrol, Austria

3 Psychiatry and Psychotherapy Clinic, University of Ulm, Ulm, Germany

**Abstract**

Correlations between rating scales are commonly interpreted as evidence of convergent or discriminant validity, yet prior studies suggest that part of these associations may be attributable to semantic similarity between item wordings rather than to genuine construct overlap alone. Building on this evidence, largely derived from bi-encoders, the present study explores whether cross-encoders, which jointly encode item pairs, offer a suitable technique for detecting semantic overlapping between questionnaire items. Using response data from the NEO-FFI and the PID5BF+M (N = 502, Labek et al., 2024), we examined whether cross-encoder-derived semantic similarity estimates are associated with empirical item correlations, and whether cross-encoders offer a systematic advantage over bi-encoders. Across twelve cross-encoder models, semantic distance was consistently negatively associated with absolute item correlations, reaching statistical significance in two-thirds of the models, with $R^2$ values of up to .37. However, cross-encoders did not consistently outperform bi-encoders based on the same base models. These findings extend prior evidence for semantic components in scale intercorrelations to cross-encoder architectures, while indicating that predictive value depends more on model-specific training characteristics than on encoder architecture itself.

## Introduction

Psychological constructs are commonly assessed using rating scales. However, several studies have indicated that associations between scale scores may, at least in part, be influenced by semantic similarities between the items (Arnulf & Larsen, 2021; Evans et al., 2022; Kjell et al. 2019; Nimon et al., 2016). This is relevant because semantically similar items may elicit comparable responses independent of genuine construct overlap, thereby contributing to observed correlations between scales. This complicates the interpretation of standard validity evidence. Correlations that are typically taken as indicators of convergent or discriminant validity (Campbell & Fiske, 1959) may partly arise from properties of item wording rather than from the constructs themselves. A key challenge, therefore, is to determine which observed relationships reflect genuine construct associations and which may be attributable to semantic similarity. Accordingly, it is important to consider approaches that allow for a more explicit examination of the role of semantic similarity, with the aim of better distinguishing between these sources of association (Labek et al., 2024).

Recent developments in Natural Language Processing (NLP) offer a way to examine this issue more directly (Demszky et al., 2023; Labek et al., 2024). Modern language models make it possible to quantify semantic similarity between texts in a systematic and reproducible manner (Devlin et al., 2019; Reimers & Gurevych, 2019). Applied to psychological questionnaires, this enables testing whether semantic relationships between items are sufficient to account for empirically observed response patterns. Previous work suggests that this is the case to some extent, as semantic similarity derived from language models has been shown to relate systematically to correlations between questionnaire items (Demszky et al., 2023; Labek et al., 2024).

However, existing approaches have primarily relied on bi-encoder architectures, in which items are encoded independently and compared at the representation level. While this design is computationally efficient, it limits the extent to which interactions between texts can be modeled. In contrast, cross-encoders jointly encode textual inputs, enabling richer interaction modeling (Reimer & Gurevych. 2019: see Figure 1). Building on this line of work, the present study examines whether cross-encoder models can capture semantic similarities between questionnaire items in a way that accounts for observed response correlations.

The analysis is based on response data from two established personality questionnaires, the NEO Five-Factor Inventory (NEO-FFI; Borkenau & Ostendorf, 1993) and the Personality Inventory for DSM-5-Brief Form Plus Modified (PID5BF+M; Bach et al., 2020). These instruments were selected because they differ in their theoretical foundations while also exhibiting well-documented conceptual overlap at the level of specific trait domains (Al-Dajani et al., 2016; Clark & Watson, 2022). This combination makes them particularly suitable for testing whether semantic similarity derived from language models aligns with empirically observed cross-instrument relationships.

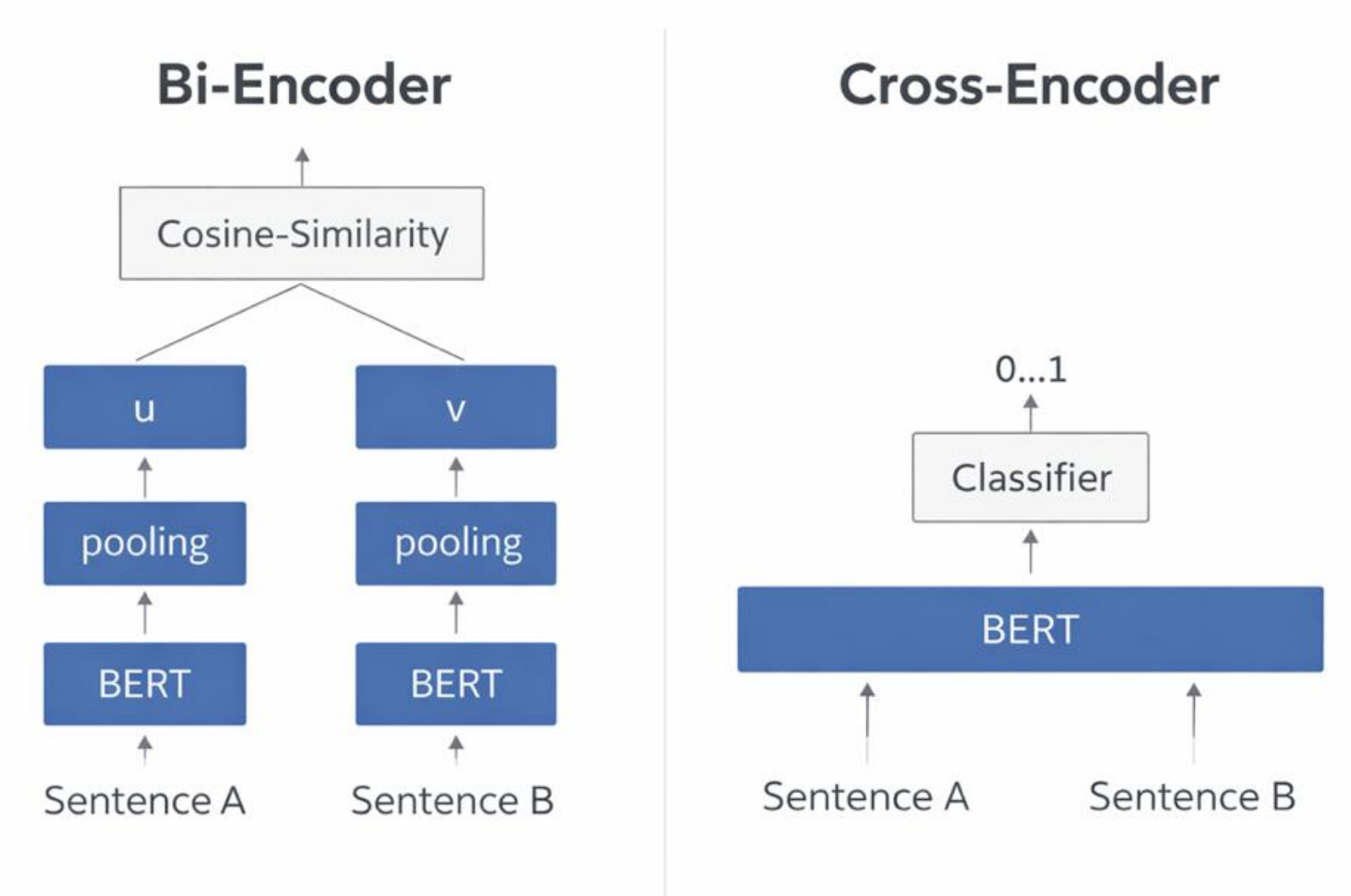

**Figure 1: Comparison of bi-encoder and cross-encoder architectures for semantic similarity estimation. In the bi-encoder (left), sentences are encoded independently into fixed representations, and similarity is computed afterward (e.g. cosine similarity). In the cross-encoder (right), both sentences are encoded jointly, enabling direct estimation of their relationship.**

Using the same dataset as Labek et al. (2024) further allows for a direct and controlled comparison with previously reported findings, ensuring methodological continuity while enabling evaluation of cross-encoder models against established bi-encoder approaches. Specifically, we investigate (a) whether semantic similarity estimates derived from cross-encoders predict empirical item correlations, (b) whether predictive accuracy is higher when item language matches model training language, and (c) whether cross-encoders provide a systematic advantage over bi-encoder approaches.

## Methods

All data were collected from 630 German-speaking healthy individuals (Labek et al., 2024). After excluding duplicates and incomplete session, a final dataset of 502 participants (mean age 24.1, std. dev. 7.78, range 22-80) was retained in the study. The sex distribution was predominantly female (376, 74.90%; 115 males, 22.90%, 7 divers 1.39%, 4 undisclosed 0.80%).

Cross-encoder similarity between all items was obtained in their German and English versions of the rating scales and retained in their original wording to preserve their semantic properties. To directly relate item semantics to response behavior, items from both instruments were systematically paired. For each item pair empirical associations were derived from response data, resulting in a complete matrix of item-level correlations.

### *Semantic Similarity Estimation*

Semantic similarity was estimated using multiple pre-trained cross-encoder models. Most models were accessed via the Hugging Face Transformers library (Hugging Face, 2025), while an additional model was accessed through the Cohere API (Cohere, 2024).

The model selection was designed to vary along key dimensions, including training objective, model size, and language coverage (Table 1). First, we included models fine-tuned for semantic textual similarity (STS), which are optimized to capture graded semantic relationships (stsb-roberta-large, stsb-roberta-base, stsb-distilroberta-base, stsb-TinyBERT-L4; Hugging Face, 2025). In addition, we included further STS-oriented model variants (ModernCE-base-sts, ModernCE-large-sts, EttinX-sts-m; Miller, 2025), which differ in their underlying base models and training configurations (Hugging Face, 2025).

To extend the analysis beyond similarity-specific training, models trained on information retrieval tasks were included (ms-marco-MiniLM-L6-v2, ms-marco-electra-base, msmarco-MiniLM-L6-en-de-v1, msmarco-MiniLM-L12-en.de-v1; Hugging Face, 2025). Finally, a multilingual reranking model (cohere rerank-multilingual-v3.0; Cohere, 2024).

**Table 1: Overview of Evaluated Cross-Encoder Models**

| Cross-Encoder | Base Model | Parameters | Training Objective | Language |
|---|---|---|---|---|
| stsb-distilroberta-base | distilbert/distilroberta-base | 82.1M | STS | EN |
| ms-marco-electra-base | google/electra-base-discriminator | ~110M | IR | EN |
| EttinX-sts-m | jhu-clsp/ettin-encoder-150m | 149M | STS | EN |
| msmarco-MiniLM-L6-en-de-v1 | microsoft/Multilingual-MiniLM-L12-H384 | ~100M | IR | EN/GER |
| ms-marco-MiniLM-L6-v2 | microsoft/MiniLM-L12-H384-uncased | 22.7M | IR | EN |
| msmarco-MiniLM-L12-en-de-v1 | microsoft/Multilingual-MiniLM-L12-H384 | ~100M | IR | EN/GER |
| ModernCE-base-sts | answerdotai/ModernBERT-base | 149M | STS | EN |
| ModernCE-large-sts | answerdotai/ModernBERT-large | 395M | STS | EN |
| rerank-multilingual-v3.0 | - | - | Reranking/IR | Multilingual |
| stsb-roberta-base | FacebookAI/roberta-base | ~125M | STS | EN |
| stsb-roberta-large | FacebookAI/roberta-large | ~355M | STS | EN |
| stsb-TinyBERT-L4 | TinyBERT (4-layer) | 14.4M | STS | EN |

Note. STS = semantic textual similarity (the model was trained to rate how similar two sentences are; IR = information retrieval (the model was trained to rank passages by relevance to a short query (e.g., on MS MARCO)); base model listed on the respective Hugging Face model card.

Importantly, the evaluated models differ in their language coverage. While several models are primarily trained on English data, others are explicitly multilingual (see Table 1). This is relevant because all questionnaire items were administered in German, allowing us to assess whether semantic similarity estimates remain predictive under potential mismatches between model training language and input language.

*Statistical Analysis*

To examine whether semantic similarity predicts empirical response patterns, we fitted models with random effects for the NEO and PID subscales to account for their repeated appearance in the pair combinations using the *nlme* package in R (Pinheiro J.C., Bates D.M, 2000, Mixed Effects Models in S and S-Plus, Springer). Z-transformed absolute empirical correlations served as the outcome variable, allowing the strength of relationships to be assessed independent of direction.

For each cross-encoder model, a separate regression model was estimated. In addition to item-level analysis, data were aggregated at the scale level by averaging semantic similarity scores and absolute correlations within combinations of NEO traits and PID domains. For these scale-level analyses, semantic distance was computed as 1 - similarity and standardized to ensure comparability across models. The aggregated data were analyzed using the same modeling approach. Model estimates included regression coefficients, standard errors, t-values, and p-values. Model fit was assessed using $R^2$ to reflect the proportion of total variance (fixed effects, random effects, and residual variance) attributable to the fixed effect of semantic distance (Nakagawa & Schielzeth, 2013).

All model inference procedures were implemented in Python (version 3.12), and statistical analyses were conducted in R (version 4.2.2).

**Results**

We first examined the associations between NEO personality traits and PID domains at the scale level. Figure 2 displays the corresponding correlations. Correlations varied in both magnitude and direction.

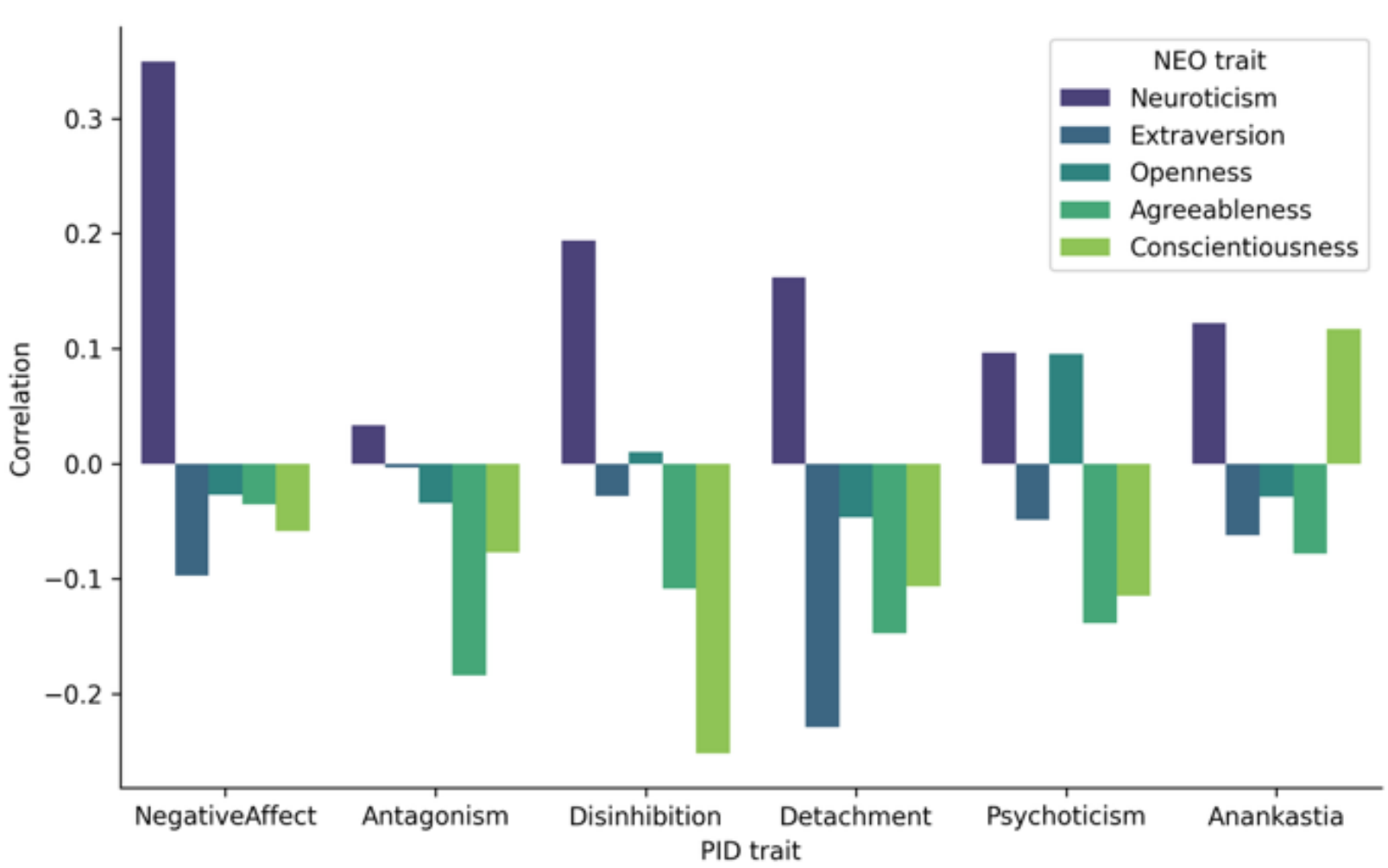

**Figure 2: Correlations between NEO personality traits and PID domains. Bars represent Pearson correlations between each NEO trait and PID domain.**

Neuroticism showed positive correlations with most PID domains, with the strongest association observed for negative affect. Smaller positive correlations were observed for disinhibition, detachment, psychoticism, and anankastia. Extraversion showed negative correlations with most PID domains, particularly with detachment. Agreeableness was negatively correlated with several domains, with the strongest negative association observed for antagonism. Conscientiousness showed negative correlations with disinhibition and detachment and a positive correlation with anankastia. Openness to experience showed generally small correlations, with a positive association with psychoticism. These patterns are broadly consistent with theoretical expectations (Al-Dajani et al., 2016; Clark & Watson, 2022) and formed the empirical basis for evaluating the predictive utility of semantic similarity estimates.

*Semantic Similarity estimates from Cross Encoders predict empirical response correlations*

To test whether semantic similarity scores derived from cross-encoder models can account for empirical response correlations, we regressed absolute item correlations on semantic distance (1 - similarity score) for each model separately. Results are summarized in Tabel 2.

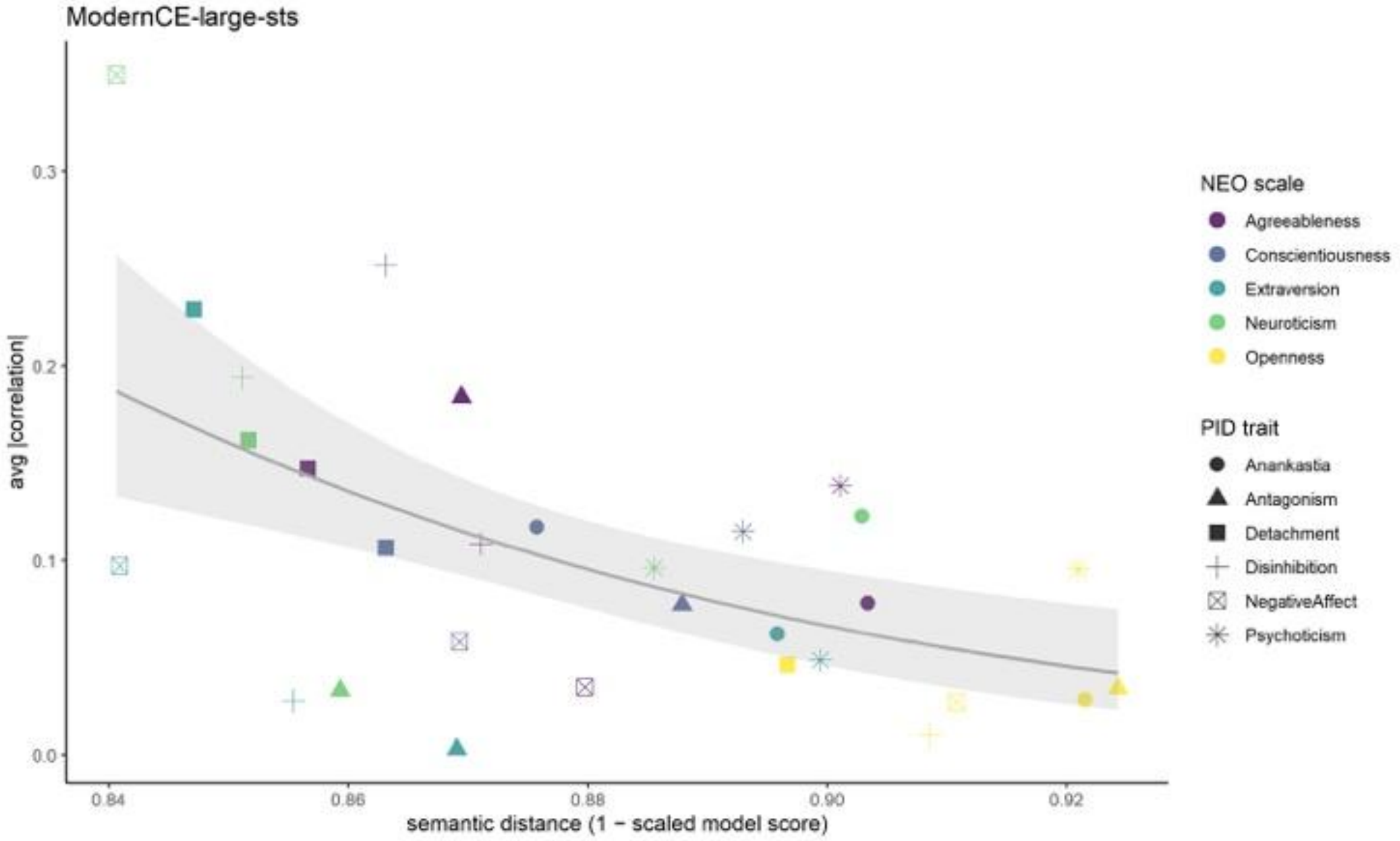

**Figure 3: Relationship between semantic distance and average absolute correlations between NEO trait and PID domains for the cross-encoder model ModernCE-large-sts. Each point represents a trait-domain pair. The x-axis shows semantic distance (1 - scaled model score), and the y-axis shows the corresponding average absolute correlation. The fitted regression line indicates a negative association, with higher semantic distance corresponding to lower absolute correlations.**

Across the 12 cross-encoder models evaluated on the German items, greater semantic distance was consistently associated with lower absolute response correlations, as reflected by uniformly negative regression coefficients. However, the magnitude and statistical significance of these associations varied considerably. For two-thirds of the models (8 out of 12), the effect reached statistical significance ($p < .05$). Among the significant models, regression coefficients ranged from $\beta = -.035$ (ModernBERT CE Base) to $\beta = -.049$ (Rerank Multilingual v3), and $R^2$ values ranged from .175 to .328, indicating moderate to strong predictive utility. The strongest model fit was observed for Rerank Multilingual v3 ($R^2 = .328$) and MiniLM-L6 (de) ($R^2 = .308$).

**Table 2: Regression Results Predicting Absolute Response Correlations from Semantic Distance Across Cross-Encoder Models, Separately for German (GER) and English (EN) Items**

| Model | Items | β | SE | t | p | $R^2$ |
|---|---|---|---|---|---|---|
| DistilRoBERTa Base | GER | -.020 | .014 | -1.35 | .187 | .055 |
| | EN | -.051 | .014 | -3.65 | .001 | .303 |
| Electra | GER | -.004 | .015 | -.27 | .787 | .002 |
| | EN | -.015 | .014 | -1.01 | .319 | .031 |
| EttinX | GER | -.039 | .014 | -2.77 | .010 | .222 |
| | EN | -.050 | .013 | -3.80 | < .001 | .336 |
| MiniLM-L6 (de) | GER | -.047 | .014 | -3.26 | .003 | .308 |
| | EN | -.041 | .013 | -3.06 | .005 | .232 |
| MiniLM-L6 v2 | GER | -.005 | .015 | -.34 | .737 | .004 |
| | EN | -.008 | .014 | -.57 | .573 | .009 |
| MiniLM-L12 (de) | GER | -.046 | .013 | -3.47 | .002 | .305 |
| | EN | -.029 | .015 | -1.98 | .058 | .125 |
| ModernBERT CE Base | GER | -.035 | .015 | -2.38 | .025 | .175 |
| | EN | -.043 | .014 | -3.15 | .004 | .250 |
| ModernBERT CE Large | GER | -.043 | .014 | -3.16 | .004 | .270 |
| | EN | -.045 | .013 | -3.42 | .002 | .276 |
| Rerank Multilingual v3 | GER | -.049 | .013 | -3.89 | < .001 | .328 |
| | EN | -.051 | .012 | -4.22 | < .001 | .370 |
| RoBERTa Base | GER | -.041 | .013 | -3.08 | .005 | .246 |
| | EN | -.042 | .014 | -3.07 | .005 | .233 |
| RoBERTa Large | GER | -.044 | .013 | -3.34 | .002 | .277 |
| | EN | -.046 | .013 | -3.52 | .002 | .285 |
| TinyBERT Base | GER | -.028 | .015 | -1.94 | .062 | .117 |
| | EN | -.034 | .014 | -2.47 | .020 | .165 |

Note. β = regression coefficient for standardized semantic distance; SE = standard error; $R^2$ = proportion of total variance (fixed + random + residual) attributable to the fixed effect (marginal $R^2$). Models were fitted as linear mixed-effects models (nlme, REML) with atanh-transformed absolute item correlations as the outcome and crossed random intercepts for NEO trait and PID domain. Significance assessed at $\alpha = .05$.

Taken together, these findings show that semantic similarity scores derived from cross-encoder models are systematically predictive of empirical response correlations, though the strength of this relationship depends substantially on the model.

*Predictive accuracy is higher when Item language matches model training language*

Given that the majority of cross-encoder models in the present evaluation were trained exclusively on English data, we hypothesized that heir predictive accuracy would be higher when applied on English rather than German item translations. To test this, the full analysis was repeated using English versions of all questionnaire items, and results were compared to the German item analyses (see Table 2).

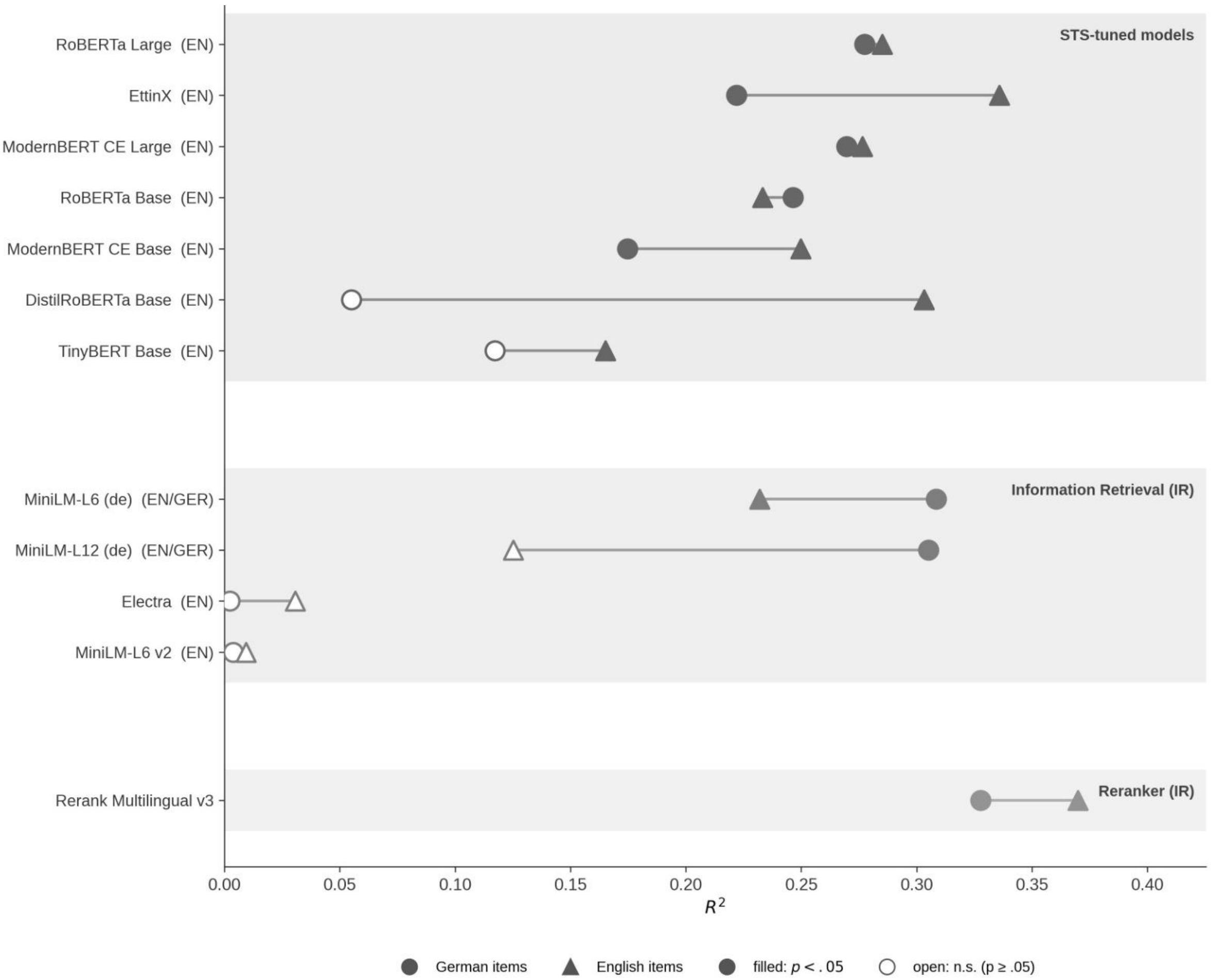

**Figure 4: R² from regressing absolute item correlations on semantic distance (mixed-effects model), shown for all 12 cross-encoders, grouped by training objective (STS, IR, Reranker). Circles = German items; triangles = English items. Filled = p < .05, open = n.s.**

For models trained exclusively on German or multilingual data, the pattern was largely consistent with a language match advantage. MiniLM-L6 (de) and MiniLM-L12 (de), both finetuned on German data, showed higher $R^2$ values for German items (.308 and .305, respectively) than for English items (.232 and .125, respectively), with MiniLM-L12 (de) showing a particularly pronounced drop in predictive accuracy for English items; for this model, the English-item effect no longer reached statistical significance (p = .058). Rerank Multilingual v3, which was trained on multilingual data, showed comparably strong performance in both languages ($R^2$ = .328 for German vs. .370 for English), consistent with its language-agnostic design.

For models trained exclusively on English data, the results were more heterogeneous. DistilRoBERTa Base showed the clearest language-match effect, with a substantially stronger result for English items ($R^2 = .303$) than for German items ($R^2 = .055$). TinyBERT Base and EttinX also showed higher $R^2$ values for English items (TinyBERT Base: .165 vs. .117; EttinX: .336 vs. .222); for TinyBERT Base, the German-item effect no longer reached statistical significance ($p = .062$). Among the remaining English-trained models, only RoBERTa Base still showed a modest advantage for German items ($R^2 = .246$ vs. .233 for English). ModernBERT CE Large and RoBERTa Large, which under the earlier model appeared to perform better on German items, showed essentially equivalent performance across languages under the mixed-effects model (ModernBERT CE Large: .270 GER vs. .276 EN; RoBERTa Large: .277 GER vs. .285 EN). ModernBERT CE Base, previously showing near-identical performance across languages, now performed noticeably better on English items ($R^2 = .250$) than on German items (.175).

Two models, Electra and MiniLM-L6 v2 showed no meaningful predictive signal in either language, with no significant results for both German and English items. This pattern suggests that their failure to predict response correlations reflects model-specific characteristics rather than a language mismatch.

Overall, models trained on German or multilingual data showed consistently higher predictive accuracy when applied to German items, consistent with a language-match advantage. Among models trained exclusively on English data, evidence for the reverse pattern was considerably weaker than initially suggested. Only RoBERTa Base retained a modest German advantage, while DistilRoBERTa Base, TinyBERT Base, and EttinX performed better on English items, and ModernBERT-CE-Base, ModernBERT-CE-Large, and RoBERTa Large showed comparable or slightly better performance on English items. These findings indicate that, once repeated measurement across NEO and PID subscales is properly accounted for, predictive accuracy for English-trained models tracks the training language more closely than previously estimated.

*Cross-Encoders provide more accurate similarity estimates than Bi-Encoders*

We hypothesized that cross-encoder models would outperform bi-encoders of the same base model, given that cross-encoders jointly encode item pairs and can therefore model interitem interactions more directly. To evaluate this, we identified comparable bi-encoder versions for three of the cross-encoder models included in the main analyses (MiniLM-L6, ModernBERT, and RoBERTa). We then repeated the regression analyses using these bi-encoder models and compared their predictive performance with that of the corresponding cross-encoders separately for German and English items (see Table 3).

**Table 3: Regression Results Comparing Cross-Encoder and Bi-Encoder Architectures, Separately for German (GER) and English (EN) Items**

| Base Model | Architecture | Items | β | SE | t | p | $R^2$ |
|---|---|---|---|---|---|---|---|
| MiniLM-L6 | Cross Encoder | GER | -.005 | .015 | -.34 | .737 | .004 |
| | | EN | -.008 | .014 | -.57 | .573 | .009 |
| | Bi Encoder | GER | -.044 | .013 | -3.29 | .003 | .277 |
| | | EN | -.059 | .011 | -5.34 | < .001 | .490 |
| ModernBERT | Cross Encoder | GER | -.043 | .014 | -3.16 | .004 | .270 |
| | | EN | -.045 | .013 | -3.42 | .002 | .276 |
| | Bi Encoder | GER | -.014 | .015 | -.94 | .354 | .028 |
| | | EN | -.053 | .012 | -4.52 | < .001 | .414 |
| RoBERTa | Cross Encoder | GER | -.044 | .013 | -3.34 | .002 | .277 |
| | | EN | -.046 | .013 | -3.52 | .002 | .285 |
| | Bi Encoder | GER | -.054 | .016 | -3.37 | .002 | .290 |
| | | EN | -.029 | .014 | -2.02 | .053 | .123 |

*Note.* β = regression coefficient; SE = standard error. Cross-encoder and bi-encoder variants share the same base model. Significance assessed at α = .05.

The results, shown in Table 3, did not support the expected pattern and instead revealed a highly model-specific pattern. For MiniLM-L6, the bi-encoder substantially outperformed the cross-encoder in both German ($R^2$ = .277 vs. .004) and English items ($R^2$ = .490 vs. .009). The MiniLM-L6 cross-encoder failed to produce a significant effect in either language, whereas the bi-encoder showed some of the strongest effects observed in the entire comparison. For RoBERTa, the pattern was more balanced. For German items, cross-encoder and bi-encoder performed comparably, with both reaching statistical significance (cross-encoder: $R^2$ = .277, p = .002; bi-encoder: $R^2$ = .290, p = .002). For English items, the cross-encoder showed a clear advantage over the bi-encoder ($R^2$ = .285 vs. .123), and the bi-encoder effect only approached, but did not reach, conventional significance (p = .053).

ModernBERT yielded the most complex pattern. For German items, the cross-encoder clearly outperformed the bi-encoder ($R^2$ = .270 vs. .028), with the bi-encoder showing a non-significant effect. For English items, however, this advantage was reversed. The bi-encoder achieved a substantially higher $R^2$ (.414) than the cross-encoder (.276). This interaction between architecture and item language within the same base model suggests that the relative utility of cross- versus bi-encoder architectures cannot be attributed to the underlying model alone, but may depend on specific combination of base model, fine tuning procedure, and the language in which similarity is estimated.

Overall, cross-encoders outperformed their bi-encoder counterparts for ModernBERT on German items and for RoBERTa on English items, but these advantages were not consistent across models or languages. For RoBERTa on German items, cross- and bi-encoder performed comparably rather than showing a clear cross-encoder advantage. In contrast, the MiniLM-L6 bi-encoder substantially outperformed the corresponding cross-encoder in both languages, and the ModernBERT bi-encoder showed superior performance for English items. These findings suggest that architectural differences alone are insufficient to explain predictive performance. Instead, the effectiveness of cross- and bi-encoder approaches appears to depend on the specific model and application context.

**Discussion**

The present study examined whether cross-encoder models can capture semantic similarities between personality questionnaire items in a way that predicts empirical response correlations, and whether they offer a systematic advantage over the bi-encoder approaches used in prior work. Overall, the findings demonstrate that semantic similarity derived from cross-encoders constitutes a meaningful predictor of item correlations, thereby extending earlier evidence obtained with bi-encoders. At the same time, predictive performance varied substantially across models. Rather than reflecting a single determining factor, this pattern of results points to an interplay of training objective, language adaption, and backbone architecture, with semantic similarity estimates varying strongly across models.

Across eight of the twelve evaluated cross-encoder models, semantic distance significantly predicted absolute item correlations for the German items, explaining up to 32.8% of the variance. These findings replicate and extend the work of Labek et al. (2024) by demonstrating that the relationship between semantic similarity and empirical response patterns is not limited to bi-encoder embeddings but also generalizes to cross-encoder architectures.

The strongest effects, however, were observed not among the STS-tuned models but among models trained for information retrieval or reranking on multilingual data, namely Rerank Multilingual v3 ($R^2$ = .328) and the German/multilingual MiniLM-L6 and MiniLM-L12 variants ($R^2$ = .308 and .305, respectively). STS-tuned models such as RoBERTa Large and ModernBERT CE Large also showed robust effects ($R^2$ = .277 and .270), consistent with the theoretical expectation that STS optimization aligns model representations with graded human similarity judgments, the construct assumed to underlie semantic overlap between questionnaire items. That the very strongest fits nonetheless emerged from IR/reranking-tuned, multilingual models suggests that exposure to diverse, cross-lingual training data may be at least as important as the specific similarity-related training objective.

The reranker family offers a particularly useful case for examining this question further, as these models share a common training objective originally developed for information retrieval (IR) by jointly encoding query-document pairs to estimate relevance (Nogueira & Cho, 2019). Despite this shared objective, their performance varied substantially depending on language coverage and fine-tuning configuration.

Specifically, MiniLM-L6 v2 (IR, en) and Electra (IR, en) failed to predict item correlations reliably whereas MiniLM-L6 (IR, en/ger), MiniLM-L12 (IR, en/ger), and Rerank Multilingual v3 exhibited some of the strongest effects in the entire model set, in several cases matching or exceeding the performance of STS-tuned cross-encoders. Overall, these results indicate that training objective alone does not fully account for the observed variability in model performance.

We further hypothesized that predictive accuracy would be systematically higher when the language of the questionnaire items matches the primary training language of the model. This expectation was largely supported for the English-trained models. DistilRoBERTa Base, TinyBERT Base, and EttinX (all STS-tuned, en) predicted English item correlations more accurately than German ones, and ModernBERT CE Large and RoBERTa Large showed essentially equivalent performance across languages rather than a reversed pattern. Only RoBERTa Base retained a modest advantage for German items despite being trained exclusively on English data. These findings indicate that language effects are not uniform across models but instead depend on the extent and nature of language-specific adaption during training.

Previous research has shown that monolingual transformer models can transfer surprisingly well to related languages even without explicit multilingual training (Artetxe et al., 2020). Such cross-lingual transfer has been attributed to shared subword structure between English and German as well as incidental multilingual exposure during pretraining (Pires et al., 2019; Blevins & Zettlemoyer, 2022). Although these mechanisms likely contribute to cross-lingual generalization, they do not adequately explain the pronounced between-model differences observed in the present study.

Having established that training characteristics account for much of the observed variability across models, the remaining question is whether encoder architecture itself contributes additional predictive value. We addressed this question by directly comparing cross-encoder and bi-encoder variants based on the same underlying base models.

We hypothesized that cross-encoders would systematically outperform bi-encoders built on the same base models due to their ability to model direct interactions between item pairs via joint encoding (Reimers & Gurevych, 2019). This expectation was not consistently supported. RoBERTa showed the expected cross-encoder advantage for English items, whereas for German items cross- and bi-encoder performed comparably; MiniLM-L6 exhibited the opposite pattern, with the bi-encoder outperforming its cross-encoder counterpart in both languages; and ModernBERT showed mixed, language-dependent effects. The inconsistent pattern may be explained by task structure. The cross-encoder advantage typically emerges in asymmetric retrieval settings where full cross-attention can exploit the imbalance between a short query and a long passage (Humeau et al., 2020), whereas questionnaire items are short and symmetric, limiting the added value of joint encoding.

The present findings demonstrate that NLP-derived semantic similarity constitutes a theoretically grounded tool for examining how linguistic item properties contribute to empirical response correlations, extending the bi-encoder approach of Labek et al. (2024) to cross-encoder architectures. In the present set of models, predictive validity appeared to track training characteristics more closely than encoder architecture per se, though this comparison was based on a small number of matched model pairs rather than a formal variance decomposition. At the same time, semantic similarity did not fully account for empirical response correlations, indicating that observed item relationship reflect both linguistic and psychological sources of covariation. From a psychometric perspective, this raises questions about construct validity. If scale correlations are partly driven by semantic overlap in item wording rather than solely by latent construct overlap, standard interpretations of convergent and discriminant validity may require qualification. Integrating computational linguistic methods into test construction and validation could therefore help clarify the extent to which observed correlations reflect latent psychological constructs rather than shared item phrasing and reduce the risk of overestimating construct-driven covariation.

### Limitations

A first limitation concerns the empirical basis of the study. The analysis is based on a single dataset comprising two personality questionnaires from one sample. The observed model rankings may therefore partly reflect dataset-specific characteristics. Second, the modeling approach. All models were evaluated in their pretrained form without task-specific fine-tuning on questionnaire items. This may have influenced performance differences across architectures and training regimes.

### Future Directions

Several directions emerge from the present findings. Replicating this study across different questionnaire types, including clinical, ability or attitude measures, would help establish the generalizability of the observed effects. Additionally, task-specific fine-tuning of cross-encoder models on psychological item pairs could yield more accurate semantic similarity estimates.

### Software availability

The code used in this study is available at https://github.com/IsabellaKainz/kainz_manuscript.

### Acknowledgments

This work was funded in part by an ERA-PERMED grant (project ArtiPro) of the FWF Austrian Science Fund (grant number I 5903) [Grant-DOI:10.55776/I5903]. ERA PerMed was supported by funding from the European Union's Horizon 2020 research and innovation programme under grant agreement No. 779282.